\documentclass[longbibliography,
 reprint,
 amsmath,amssymb,
 aps,
 prx,
]{revtex4-2}

\usepackage{graphicx}% Include figure files
\usepackage{dcolumn}% Align table columns on decimal point
\usepackage{bm}% bold math
\usepackage{braket}

\begin{document}

\preprint{APS/123-QED}

\title{Topological Defect Propagation to Classify Knitted Fabrics}

\author{Daisuke S. Shimamoto}\email{shimamoto-daisuke806@g.ecc.u-tokyo.ac.jp}
\affiliation{%
Graduate School of Arts and Sciences, The University of Tokyo, 3-8-1 Komaba, Meguro, Tokyo, 153-8902, Japan
}

\author{Keiko Shimamoto}
\affiliation{Independent researcher, Tokyo, Japan}

\author{Sonia Mahmoudi}
\affiliation{Advanced Institute for Materials Research, Tohoku University, 2-1-1 Katahira, Aoba-ku, Sendai, 980-8577, Japan}

\affiliation{RIKEN iTHEMS, 2-1 Hirosawa, Wako, Saitama, 351-0198, Japan}

\author{Samuel Poincloux}
\affiliation{%
Department of Physical Sciences, Aoyama Gakuin University, 5-10-1 Fuchinobe, Sagamihara, Kanagawa 252-5258, Japan
}%

\date{\today}

\begin{abstract}
Knits and crochets are mechanical metamaterials with a long history and can typically be produced from a single yarn. 
Despite the simplicity of the manufacturing process, they exhibit a wide range of structural configurations with diverse mechanical properties and application potential. 
Although there has been recent growing interest in textile-based metamaterials, a rigorous topological characterization of what makes a structure \textit{knittable} has been lacking. 
In this paper, we introduce a general criterion based on topological constraints that distinguishes knits and crochets from other textile structures. 
We demonstrate how the introduction of topological defects and their propagation makes this classification practical. 
Our approach highlights a fundamental link between manufacturing processes and structural fragility. 
Within this framework, we show how the rationalization of defect propagation unlocks the design of fabrics with controllable resistance to damage.
\end{abstract}

\maketitle

\section{Introduction}

Textiles are among the oldest human technologies and continue to evolve as both traditional crafts and advanced manufacturing techniques~\cite{emery1995primary,seiler1994textiles,good2001archaeological,friedman2023joachim}. They include a wide variety of structures produced through processes such as weaving, knitting, crochet, braiding, and non-woven fabrication, each created through distinct methods of intertwining yarns in space. 

Fundamentally, textiles are formed by intertwining yarns through repeated local operations. 
The resulting structures can be modeled as planar assemblies of one-dimensional curves that pass over and under each other. 
This dimensional reduction, from a three-dimensional physical structure to a two-dimensional diagram, makes it possible to study the textiles using tools from knot theory~\cite{grishanov2009topologicalI,grishanov2009topologicalII,mahmoudi2020classification,markande2020knotty,fukuda2023classification, PAFDC2025}. 
Knot theory is a branch of mathematics concerned with the classification and properties of entangled curves in three-dimensional space, often studied through their two-dimensional projections called knot diagrams~\cite{adams2004knot}.
The topology of a fabric, specifically the sequence and arrangement of yarn crossings, is known to have profound effects on its physical properties~\cite{knittel2020modelling,storck2022topology,singal2024programming,niu2025geometric}. 

Among textiles, knits and crochets form a special category in which a single yarn can be used to create a self-supporting fabric through local loop operations. 
These loop-based fabrics exhibit properties that go beyond the behavior of the yarn itself. 
While the yarns are typically inextensible, the resulting fabrics display remarkable stretchiness. 
Knits and crochets then possess the characteristic features of metamaterials whose macroscopic properties arise not only from composition, but also from geometry and topology.

In particular, the arrangement of crossings is shown to largely determine the mechanical response of the fabric~\cite{singal2024programming}. Recent studies have also investigated their response to deformation~\cite{liu2017role,poincloux2018geometry,singal2024programming,gonzalez2024pulling}, mechanical stability~\cite{crassous2024metastability}, global shape~\cite{belcastro2009every,chas2018crochet}, and curvature emergence~\cite{kurbak2008basic,tajiri2025curling,niu2025geometric}, reflecting the complex interplay between geometry, topology, and material behavior. 

Although knitting and crochet differ in both construction process and physical properties~\cite{karp2018defining}, they share essential topological features: both involve the formation of looped entanglements along a continuous yarn. 
To emphasize this common structure, we refer to them collectively as \textit{amimono}, adopting the Japanese term that encompasses both techniques. 
To distinguish amimono from other textiles, here we establish a direct link between the way a textile can be manufactured and the topology of its entanglements.

Despite recent advances in modeling textiles, a general topological definition of what makes a textile realizable as an amimono, referred to as \textit{knittable} in this paper, remains unclear. 
As used in this work, the term amimono refers to a broad class of loop-based textile structures that we consider knittable, including warp knitting, flat and circular weft knitting, crochet, and lace.
Most research has focused on standard patterns, such as jersey knitting and its common variants, leaving the broader classification problem open. Attempts to create textiles with more complex topologies, and thus to define and determine what knittability is from a topological viewpoint, are surprisingly difficult~\cite{rutt1987history}.

In this study, we propose a mathematical framework rooted in knot theory for identifying the topological properties that define knittable textiles. 
Starting from a textile diagram, we develop a method to translate the question of knittability into determining the triviality of a knot or a link. 
In particular, we evaluate how the introduction of topological defects in a textile diagram, as well as their propagation, can disentangle the structure.
This classification reveals qualitative differences between amimono and other textiles, and further distinguishes between knits and crochets based on how these defects behave within the structure.

\section{Knitting process and textile classification \label{SETUP}}

\begin{figure}[b]
    \centering
    \includegraphics[width=8.6cm, bb = 0 0 261 93]{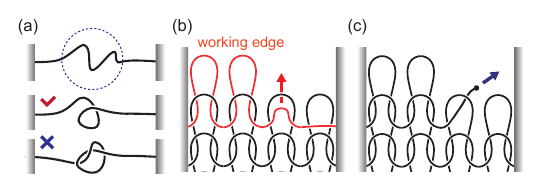}
    \caption{(a) Configurations that can and cannot be created from the initial state shown on top, using only local manipulations within the dashed circle.
    (b) Operations that create or remove crossings without moving the yarn ends are permitted in the process of creating a knitting.
    (c) Operations that move the end points of the yarn are not permitted.}
    \label{operation}
\end{figure}

We model textiles as quasi-two-dimensional structures composed of intertwined one-dimensional yarns. 
This broad definition allows an infinity of possible patterns, which are mainly determined by the crafting or industrial technique used to produce them.
Knitted and crocheted textiles, commonly referred to as amimono fabrics, result from a very specific process where only local manipulation of the yarn is permitted while its two ends are pinned, see Fig.~\ref{operation}(a).
Amimono fabrics essentially consist of sliding loops, which refers to a part of the yarn forming an omega shape, into previous ones (Fig.~\ref{operation}(b)), whereas threading the yarn into a previous loop is forbidden (Fig.~\ref{operation}(c)). 
Knitting and crocheting techniques can be modeled on a grid where vertically entangled loops form stitches along horizontal yarn passes. They differ by the order in which the stitches are created. 
These seemingly restrictive conditions permit the production of amimono fabrics from a single yarn, with one end free at the start, and the other end still deeply buried in a ball or spool of yarn. This remarkable feature of amimono has contributed to its popularity, as knitters only need a yarn ball and small tools (such as needles or hooks) to manipulate the yarn locally. Since all materials can be easily carried, amimono fabrics can be made anywhere by anyone, outside of a workshop, and without the need for expensive machinery.
Nevertheless, highly sophisticated and innovative materials can be created from these simple processes, including crocheting the hyperbolic plane~\cite{taimina2018crocheting}, artificial muscles~\cite{maziz2017knitting}, and actuators~\cite{abel2013hierarchical}, among others.

\begin{figure}
    \centering
    \includegraphics[width=8.6cm, bb = 0 0 261 261]{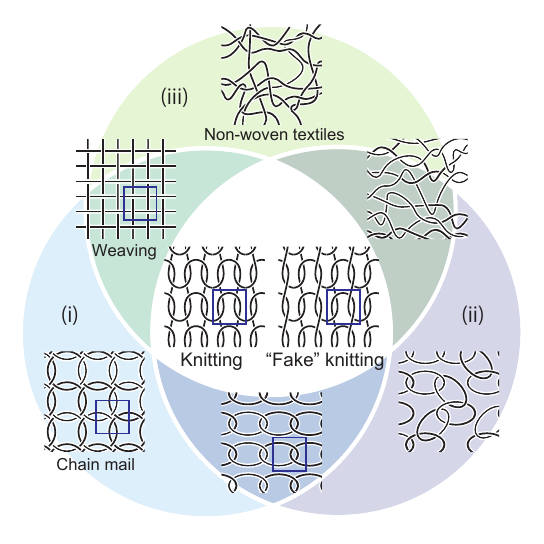}
    \caption{Venn diagram to classify textiles, built from the three conditions detailed in the text. 
    Each intersecting region features an example of a textile satisfying the corresponding conditions.
    The imposed conditions rule out textile families that are clearly distinct from knits. 
    Our focus is on the center of the Venn diagram, with subtle differences between amimono (such as a Jersey knit, center left) and `fake' knitting (center right). 
    For each textile with discrete translational symmetry, a possible unit cell is depicted by a blue square.}
    \label{venn}
\end{figure}

Another consequence of this local fabrication process is the presence of a working edge where loops are added gradually. By working edge, we mean the yarn that creates the highest row of stitches, considering a bottom-to-top construction of the amimono fabric, as illustrated in Fig.~\ref{operation}(b). 
Local manipulations of the yarn to create the fabric are reversible along this edge, which then constitutes a ``weak point'' of the fabric. An amimono fabric can then be characterized as a textile that can be untangled only via its working edge, while the bulk is topologically protected (excluding damaged fabrics). 
This topological protection and the sensitivity to defect propagation from the working edge will be at the core of our knittability criteria.
However, from a textile pattern alone, distinguishing structures fabricated via an amimono process can be far from obvious (see ``real'' and ``fake'' knitting in the center of Fig.~\ref{venn}). 
Still, some other classes of textiles, such as chain-mail or weaves, can be easily ruled out. The method we develop to assess knittability will focus on patterns that are not easily distinguishable.

Translating the physical characteristics of knits and crochets into mathematical conditions is our first step in assessing if a fabric can be made by one of these two techniques. 
To distinguish amimono fabrics from other textiles, we first impose the following conditions for the pattern:

\begin{itemize}
    \item[(i)] discrete periodicity in two transverse directions, 
    \item[(ii)] infinitely many yarns running in parallel directions, such that each yarn crosses exactly two other yarns,
    \item[(iii)] no closed yarns.
\end{itemize}

The first periodicity condition (i) rules out non-woven textiles that are made by a random arrangement of a multitude of yarns. This condition also means that we restrict ourselves to swatches of fabric with homogeneous patterns, and the knittability criteria will reflect a local property of the structure, independent of its global shape. 
The third condition (ii) aims to eliminate woven-like fabrics where threads run in at least two distinct directions, as well as braided-type structures. This condition, by possibly ruling out some complex knit and crochet patterns, restricts the applicability of our method, but significantly simplifies its demonstration. 
Note that even though amimono can be made from a single yarn, we focus here on the bulk properties of fabrics, ignoring its edges. The successive passes of the same yarn are then equivalent to multiple parallel yarns. 
Condition (iii) eliminates chain-mail-like patterns featuring closed loops, which are readily incompatible with knitting manufacturing methods using open yarns.
In Fig.~\ref{venn}, the different conditions and examples of associated patterns are presented in a Venn diagram. 
The structures of interest in the study are those satisfying the three conditions at the center of the diagram. 
The three conditions leave ample room for unknittable diagrams, such as the "Fake" knitting on the center right of Fig.~\ref{venn}, that is distinct from the classic Jersey knit patterns on its left. 
To distinguish how these subtle changes in textile diagrams lead to dramatic differences in manufacturability, we use tools from knot theory.

\section{Topological model of doubly periodic textiles}

The key steps and concepts of this paper will be presented at two levels of comprehension: one in plain language, the other in more rigorous mathematical terms. This section introduces the tools and definitions from knot theory that are necessary to follow the reasoning in plain language. The mathematical description requires more advanced concepts that interested readers may find in~\cite{adams2004knot}.

\begin{figure}
    \centering
    \includegraphics[width=8.6cm, bb = 0 0 503 308]{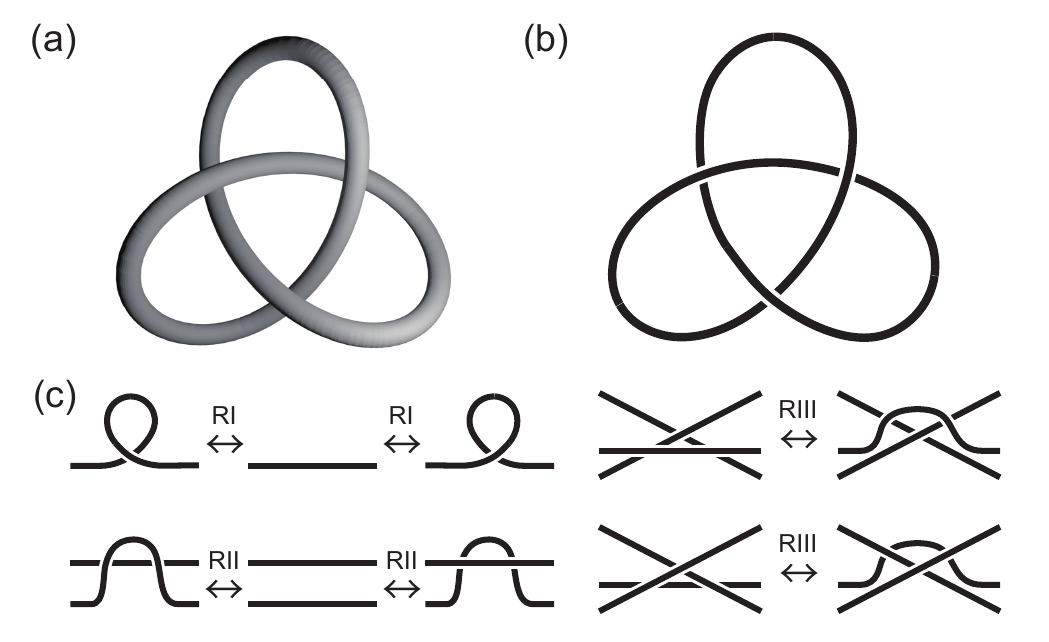}
    \caption{(a) Illustration of a trefoil knot in 3D space, and (b) its corresponding knot diagram. 
    (c) Illustration of the three types of Reidemeister moves, RI, RII, and RIII.
    }
    \label{knotdiagram}
\end{figure}

Knot theory is the study of closed, entangled curves in space, called links (Fig.~\ref{knotdiagram}(a)). In particular, a link consisting of a single curve component is referred to as a knot.
Two links are said to be ambient isotopic, or simply equivalent, if one can be transformed into the other by continuous deformations without cutting or gluing the curves.
Evaluating their equivalence in three-dimensional space is not trivial.
Knot theory reduces the complex 3D equivalence problem to a simpler 2D combinatorial problem through the introduction of link diagrams and isotopy moves.
A link diagram is a two-dimensional projection of a link (Fig.~\ref{knotdiagram}(b)) in which over/under crossing information is marked at each intersection between two arcs.
The continuous deformations of links are then discretized into a set of local transformations on their planar diagrams, called Reidemeister moves, which capture three-dimensional equivalence. 
There are exactly three types of Reidemeister moves (Fig.~\ref{knotdiagram}(c)):
\begin{itemize}
    \item[RI:] Add or remove a single loop.
    \item[RII:] Add or remove two crossings.
    \item[RIII:] Slide a strand past a crossing.
\end{itemize}
Two links are equivalent if and only if their diagrams are related by a finite sequence of Reidemeister moves. This statement is known as the Reidemeister Theorem. 
In particular, we call a knot trivial (or an unknot) if its diagram can be transformed into a simple circle without any crossings by Reidemeister moves. 
A trivial link is a set of disjoint unknots, meaning that its corresponding diagram contains no crossings, up to Reidemeister moves.

\begin{figure}
    \centering
    \includegraphics[width=8.6cm, bb = 0 0 503 359]{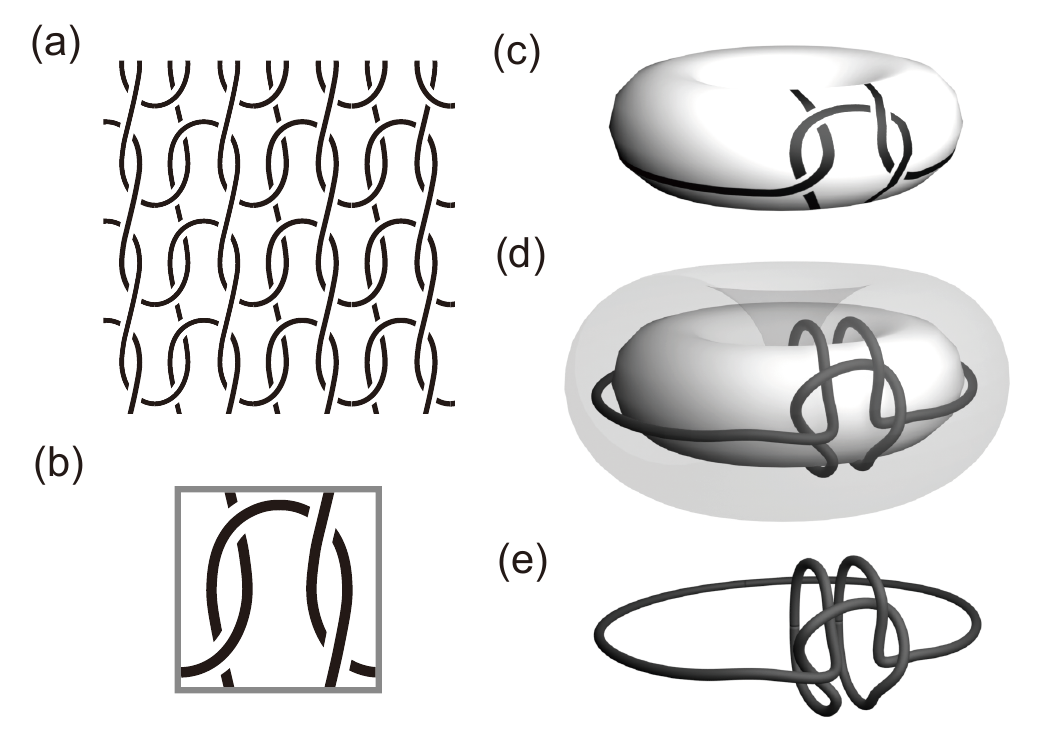}
    \caption{(a) A 2D textile diagram in the plane. 
    (b) A unit cell extracted from the textile diagram. 
    (c) A unit cell folded on a torus. 
    (d) A knot generated by lifting the diagram from the torus to a thickened torus. 
    (e) A knot in three space obtained by forgetting the thickened torus in three space. The resulting knot is a trefoil knot, which, when projected onto a plane, has a knot diagram equivalent to Fig.~\ref{knotdiagram}(b).}
    \label{torus}
\end{figure}

In this paper, we consider doubly periodic textile diagrams that extend infinitely in the Euclidean plane $\mathbb{R}^2$ (condition (i)), such as the one shown in Fig.~\ref{torus}(a). 
Intuitively, a simple cutting and stitching process permits one to build a link diagram from a textile one. We first extract a parallelogram unit cell from the textile pattern (Fig.~\ref{torus}(b)), then glue its opposite sides together to form a toroidal shape (Fig.~\ref{torus}(c)). 
The double periodicity ensures that when gluing the opposite sides of the unit cell, each cut curve can be joined continuously with another one from the other side. By interpreting back the diagram on the torus as a 3D curve (Fig.~\ref{torus}(d)), and removing the torus, we obtain a 3D knot (Fig.~\ref{torus}(e)).

More formally, the object resulting from the gluing is a link diagram on the torus $T^2$, which we refer to as a torus diagram. 
A torus diagram is obtained by the quotient of a doubly periodic textile diagram under a periodic lattice (see~\cite{mahmoudi2020classification,kawauchi2018complexities,bright2020encoding} for more details). 
In this paper, we use the term unit cell to refer to a torus diagram cut along a longitude and a meridian of the torus.
A torus diagram can then be lifted in 3D to a set of entangled curves in the thickened torus $T^2\times I$, with $I$ being the unit interval (Fig.~\ref{torus}(d)). 
By viewing the thickened torus in the three-space $\mathbb{R}^3$, one can obtain a link in $\mathbb{R}^3$ by ``forgetting'' $T^2\times I$, as illustrated in the example of Fig.~\ref{torus}(e), which can be regarded as a trefoil knot in $\mathbb{R}^3$.
However, the resulting link depends on the choice of the unit cell.
Rigorous mathematical investigations as presented in~\cite{grishanov2009topologicalI,grishanov2009topologicalII,markande2020knotty,diamantis2023equivalence}, establish the equivalence between doubly periodic textile structures in three-space through the torus diagrams associated to their unit cells.
The definition of a suitable unit cell is then a crucial step in applying knot theory to textile diagrams.

\section{Unit cell\label{UNIT}}

In analogy to crystals, modeling textile structures as doubly periodic planar diagrams allows the extraction of unit cells, which reduces the complexity of the problem. 
A unit cell can be viewed as a portion of the textile diagram that can be tiled infinitely in two transverse directions. 
For any doubly periodic diagram, an 
infinite number of unit cells exists and the choice of a specific one is arbitrary.
Knittability is deeply rooted in the knitting and crocheting processes of creating iteratively new loops along a line or a column. 
A natural representation of an amimono then takes the form of a matrix, with one element representing one unit cell, and the fabrication process consists of adding a new line on top of the matrix.  The yarn then travels along the lines, and the loops are connected along columns.

\begin{figure}
    \centering
    \includegraphics[width=8.6cm, bb = 0 0 261 261]{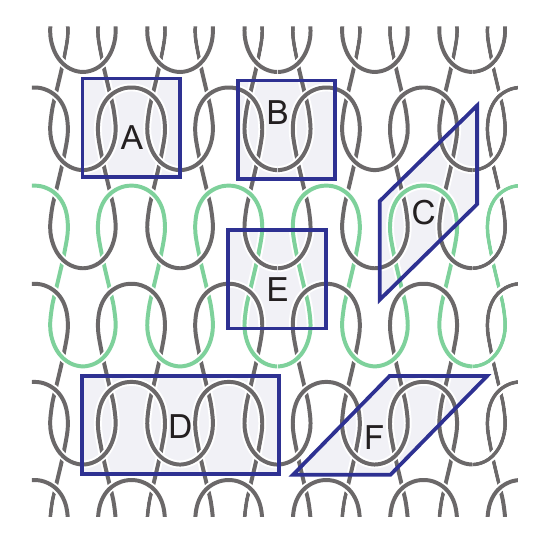}
    \caption{A-F: Examples of possible unit cells for the jersey knit pattern. 
    Only cells A and B satisfy all our conditions for a unit cell. 
    Condition (I) is not respected by cells C and F, while unit cells C and E violate condition (II). 
    Unit cell D is only not minimal, hence not satisfying (III).
    }
    \label{unitcell}
\end{figure}

In the following, we introduce the notations and conditions to rigorously select a suitable unit cell for our classification method.
Our knittability assessment methodology will rely on a selected group of unit cells that satisfy specific conditions, chosen to reflect the amimono process.
A yarn pass refers to the continuous path of a yarn running through the fabric (Fig.~\ref{unitcell}). Each unit cell can be represented as a parallelogram in a lattice, bounded by four edges. 
The edges that are non-parallel to the overall yarn pass direction are called vertical edges. The conditions then read:

\begin{itemize}
    \item[(I)] Unit cells are chosen so that all yarns passing through the vertical edges of the unit cell are of the same pass.
    \item[(II)] Unit cells are chosen so that they encompass only two passes of yarn.
    \item[(III)] The unit cell is taken to be the smallest in area.
\end{itemize}

By isolating two yarn passes, conditions (I) and (II) will be essential for our methodology and the definition of topological defects. 
It also simplifies the representation of an amimono diagram as a matrix with each line corresponding to one yarn pass. 
These conditions exclude some textile patterns that cannot be reduced to unit cells with only two passes. 
While extensions of our method to more general cases is possible, in this paper we will focus on patterns satisfying these conditions for simplicity.
Condition (III) is implemented to reduce the number of possible unit cells, and also the number of topological operations needed further down to assess knittability. 
Defining a minimal unit cell helps as well to compare coherently different patterns, when assessing their complexity for instance. 
Examples of unit cells satisfying or not these conditions are illustrated in Fig.~\ref{unitcell}.

With unit cells and textile patterns satisfying these conditions, a patch of the diagram of our textiles of interest can be represented simply in matrix form. The position of each unit cell is defined by a pair of integers $(i,j)$, with an arbitrary reference $(0,0)$ on the bottom left of the matrix.
A unit cell located directly above the cell at position $(i,j)$, across a horizontal edge, is assigned the position $(i+1,j)$. 
Similarly, a unit cell located directly to the right of the cell at position $(i,j)$, across a vertical edge, is assigned the position $(i,j+1)$.  
For each fixed $i$, the set of unit cells with position $(i,j)$ for all $j$ is parallel to what we refer to as the $i$-th pass of the yarn.

\section{Definition and propagation of defects\label{defect_propagation}}

\begin{figure}
    \centering
    \includegraphics[width=8.6cm, bb = 0 0 251 410]{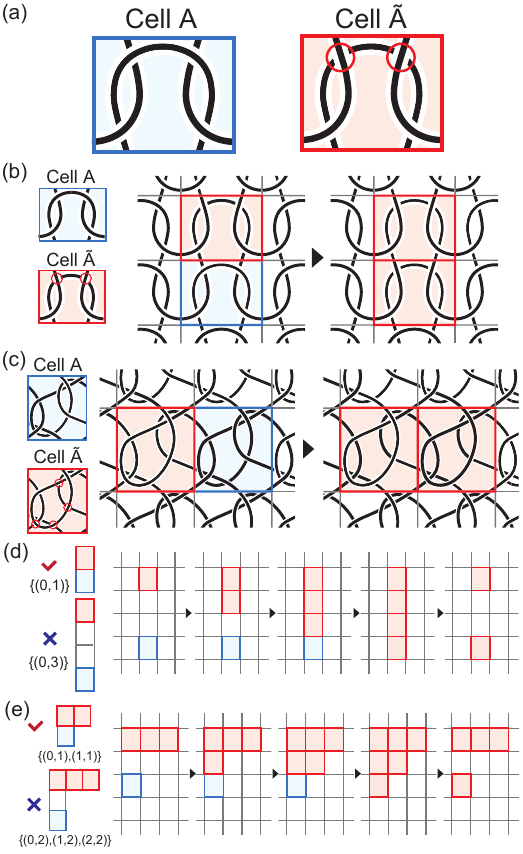}
    \caption{
    (a) Illustration of a unit cell and its defect counterpart. The crossings that have been switched are circled in red. (b-c) Propagation of defects in a Jersey (b) and crochet (c) pattern. 
    The respective propagation sets are $\{(1,0)\}$ and $\{(0,-1)\}$. 
    (d) Illustration of the reduction of the propagation set $\{(3,0)\}$ to $\{(1,0)\}$ by applying several times the latter. 
    The cell $A_0$ (blue) can be transformed into a defect and the cells $(0,1)$ and $(0,2)$ can be recovered to the initial state by applying the propagation pattern corresponding to the propagation set $\{(0,1)\}$ if the cell $(0,3)$ is $\tilde A$. 
    (e) Likewise for the propagation sets $\{(2,0),(2,1),(2,2)\}$ and $\{(1,0),(1,1)\}$.
    }
    \label{copy}
\end{figure}

The conditions we impose on textile diagrams and the choice of their unit cells allow for a simplified view of their underlying structure.
A finite subset of the diagram is represented as a rectangular matrix of repeated unit cells. 
The same yarn runs in and out of the lateral sides of a unit cell, and hence along a line of the matrix. 
A remarkable feature of amimono is the contrast between the topological protection of their bulk and the fragility of their edges. 
To capture this feature, we define a defect alter ego of a unit cell and describe how these topological defects can propagate to adjacent cells.
We transform a unit cell $A$ into its corresponding defect $\tilde A$ by disentangling the yarns passing through it. 
To disentangle, we switch the crossings of the yarns belonging to different passes, such that a yarn at line $i$ is always under the next yarn on line $i+1$ (Fig.~\ref{copy}(a)).

A rigorous approach is proposed as follows.
Considering a unit cell $A$ in the $i$-th line of the grid, we define its corresponding defect unit cell $\tilde A$ as the least tangled version of $A$. 
To do so, as in classical knot theory~\cite{adams2004knot} and by using a similar approach to the one presented in~\cite{toky} for the case of triply periodic structures, we use the concept of unknotting operations to transform $A$ locally. 
An unknotting operation consists of switching the over/under crossing information at a crossing of $A$ as illustrated in Fig.~\ref{copy}(a). 
Then, to obtain the least tangled version of $A$, we apply unknotting operations to transform $A$ into a unit cell, reaching the minimum number of crossings possible up to Reidemeister moves, known as the crossing number. 
More precisely, we first select an arbitrary crossing of $A$ to be switched, and denote by $A_1$ the resulting diagram. 
We repeat the process to obtain $A_2$, which differs from $A_1$ by a single unknotting operation, and so on until reaching a unit cell $A_n$ whose crossing number is the least among all the unit cells $\{A_1, A_2, \cdots, A_n\}$, connected by unknotting operations.
This process generates the defect unit cell $\tilde A$ from $A$. This approach can be computationally expensive for unit cells containing a large number of crossings.
We now present a practical approach to construct the defect unit cell $\tilde A$ from $A$ that captures the actual process of creating knits and crochets. Recall that these amimono fabrics are created from untangled yarns. We can thus model a yarn as a line $Y=\mathbb{R}$ in the $x$-axis of the three-space $\mathbb{R}^3$ and assign to each point $\alpha$ of $Y$ a coordinate $x_{\alpha}$ in $\mathbb{R}$. Then, an amimono structure can be seen as the image of $Y$ under a map encoding the construction process of the fabric. This 3D fabric can then be projected as a diagram $D$ on a grid in $\mathbb{R}^2$, as presented earlier. A point $\alpha$ in $Y$ is mapped to a point $\alpha'$ with coordinates $(x'_{\alpha}, y'_{\alpha})$ in $D$. Consider now two points $\alpha$ and $\beta$ in $Y$ which map to the same crossing in a unit cell $A$ in $D$. This implies that $x'_{\alpha} = x'_{\beta}$ and $y'_{\alpha} = y'_{\beta}$. Then, if $x_{\alpha} \leq x_{\beta}$ in $Y$, the crossing is switched in $A$, otherwise, it remains unchanged. This process also generates a defect unit cell $\tilde A$, which is a disentangled version of $A$. 

A defect unit cell $\tilde A$ is thus created by locally disentangling one thread in a unit cell $A$. 
Introducing defects into a textile diagram modifies its local topology. 
We now assess whether the topological defects remain localized or propagate to neighboring unit cells. 
To assess the propagation of a defect, we place several defects $\tilde A$ in a given grid of unit cells $A$. 
Then, by evaluating the topological equivalence through Reidemeister moves, it is possible to assess if the original neighboring unit cells $A$ can be transformed into defects $\tilde A$, as shown in Fig.~\ref{copy} (b,c).
In the first example of Fig.~\ref{copy}(b), the cell $\tilde A$ can propagate to adjacent cells perpendicular to the direction of the yarn, relative to the original defect. 
While in the second example in Fig.~\ref{copy}(c), the cell $\tilde A$ can propagate to an adjacent cell parallel to the direction of the yarn. 
The sequences of Reidemester moves allowing the propagation of defects in these two examples are detailed in the Appendix.
Thus, the defects do not necessarily propagate along the yarn, but can be propagated in various directions, depending on the topology of the cell $A$ and the cell $\tilde A$. 
We also observe that numerous topologies do not allow the propagation of a defect.

We now introduce a formalization and classification of the rules of defect propagation.
Select first an arbitrary unit cell, denoted $A_0$, in the matrix representing the textile diagram of interest.
Our objective is to evaluate the number and the position of the neighboring unit cells of $A_0$ which need to be changed into $\tilde A$, so that $A_0$ can be transformed into $\tilde A$ by Reidemeister moves. 
A \textit{propagation set} $\mathcal{S}$ will then be defined as a set of unit cell relative positions, which does not include $A_0$.
More precisely, we first consider the set of all unit cells in the matrix, identified by their relative position from $A_0$, $\mathcal{P} = \{(i_1,j_1), (i_2,j_2), \cdots\}$.
The propagation set $\mathcal{S}$ is then defined as a subset of $\mathcal{P}$ satisfying the following propagation conditions:

\begin{itemize}
    \item[(1)] 
    If all the unit cells whose positions belong to $\mathcal{S}$ are transformed into defect cells, then there exists a sequence of Reidemester moves that transform the cell $A_0$ into a defect $\tilde A$, while keeping the other cells unchanged. 
    We then refer to a \textit{propagation pattern} the transformation of $A_0$ to $\tilde A$, or conversely, from $\tilde A$ to $A_0$. 
    This condition and resulting propagation pattern are independent of the type of unit cells ($A$ or $\tilde A$), whose positions do not belong to $\mathcal{S}$.
    
    \item[(2)] 
    Propagation sets that can be reproduced by repeating other propagation patterns are excluded.
    
\end{itemize}

Propagation condition (1) sets the requirement for defects to propagate. 
If for the set of unit cell relative positions $\mathcal{P}$, there is no subset $\mathcal{S}$ that satisfies condition (1), then the pattern does not allow defect propagation. 
The Jersey knit pattern (Fig.~\ref{copy}(b)) has a propagation set $\{(1,0)\}$, so only one defect just above this position is sufficient to transform $A_0$ into a defect $\tilde A$. 
In contrast, with a crochet-type pattern (Fig.~\ref{copy}(c)), defects propagate with a propagation set $\{(0,-1)\}$, hence a single defect on the left of $A_0$ is enough to transform it into $\tilde A$ by Reidemeister moves.
Condition (2) reduces the propagation set $\mathcal{S}$ to a minimal set. This condition eliminates the propagation sets that are equivalent to the repetition of propagation patterns corresponding to other propagation sets. 
In the case of the jersey (Fig.~\ref{copy}(b)), the propagation set $\{(3,0)\}$ also satisfies condition (1) as defects can propagate on the whole column below a defect. 
However, it does not satisfy condition (2) as the propagation set $\{(3,0)\}$ can be obtained by repeating the propagation set $\{(1,0)\}$, as illustrated in Fig.~\ref{copy}(d). 
Another example is shown in Fig.~\ref{copy}(e), where the propagation set $\{(2,0),(2,1),(2,2)\}$ can be reduced to the propagation set $\{(1,0),(1,1)\}$ by applying it multiple times. 
We can note that once a propagation set has been established, the topology of the textile pattern is no longer needed to study the propagation of defects and the equivalence between propagation sets. 
In the following, we will utilize the concept of topological defects to define knittability criteria, and then apply the propagation set to provide a straightforward method for implementing these criteria.

\section{Knittability criteria}

We propose two topological constructions that define criteria for knittability and classify amimono. 
These constructions and ensuing criteria rest on the remarkable property that amimono can be untangled from its working edge only.
Here, the working edge (Fig.~\ref{operation}(b)) can be characterized in a unit cell as a strand that does not belong to the $i$-th pass and that is not tangled with the $i$-th pass.

\begin{figure*}
    \centering
    \includegraphics[width=15cm, bb = 0 0 399 403]{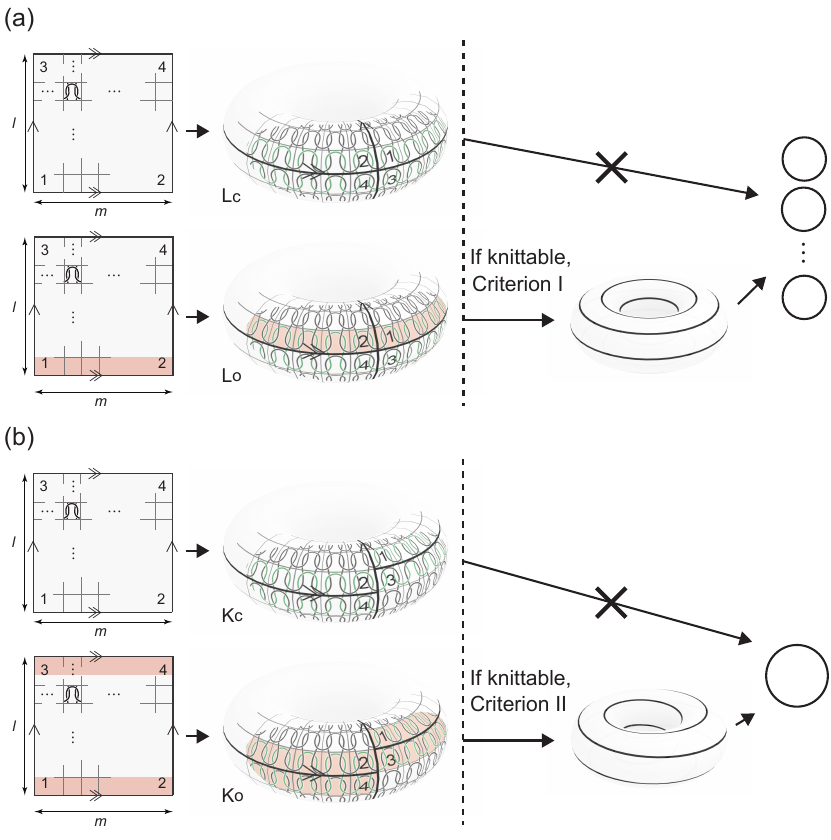}
    \caption{Illustrations of the two knitting criteria. (a) Criterion I: A $m\times l$ patch $G_{\rm c}$ is glued into a torus to form a link $L_{\rm c}$. We introduce a row of defects in $G_{\rm c}$ and fold the resulting grid $G_{\rm o}$ into the link $L_{\rm o}$. If the 3D versions $L'_{\rm c}$ is not a trivial link but $L'_{\rm o}$ simplifies to a set of disjoint loops, the criterion I is satisfied.
    (b) Criterion II: Here, $G_{\rm o}$ is defined by transforming 2 consecutive rows into defects, and then the knots $K_{\rm c}$ and $K_{\rm o}$ are obtained by folding $G_{\rm c}$ and $G_{\rm o}$ onto a torus but shifting the lateral sides by one row. If the 3D knot $K'_{\rm c}$ is not trivial but $K'_{\rm o}$ simplifies to a simple loop, criterion II is satisfied.
    }
    \label{criteria}
\end{figure*}

In both criteria we will introduce, we start from a doubly periodic textile diagram generated by a unit cell $A$.
Next, we consider a finite subset $G_{\rm c}$ of this diagram, consisting of $l$ rows and 
$m$ columns of unit cells $A$ arranged in an $(l,m)$-grid.
We then replace $n$ lines of $G_{\rm c}$ by $n$ lines of defects $\tilde A$\, and we denote by $G_{\rm o}^n$ this new diagram. The disentanglement of threads induced by the defects effectively introduces a working edge in $G_{\rm o}^n$. 
Further, by definition of $G_{\rm c}$ and $G_{\rm o}^n$, we can identify the opposite sides of the rectangular grid and obtain a torus diagram.
The knittability criteria we now introduce are based on the comparison of the topological properties of the links in three-space that can be created from the grids $G_{\rm c}$ and $G_{\rm o}^n$.
In a word, knittability has the consequence of producing trivial links if folded from the subset with a working edge $G_{\rm o}^n$, while the links folded from $G_{\rm c}$ remain topologically protected. The topological processes corresponding to the two criteria are illustrated in Fig~\ref{criteria}.

Knitting Criterion I --- 
We construct a link in the three-space $\mathbb{R}^3$ from a grid $G_{\rm c}$, made of unit cells $A$, following the idea illustrated in Fig.~\ref{torus}. 
First, since $G_{\rm c}$ forms a torus diagram by gluing its opposite sides, we can obtain a link $L_{\rm c}$ in the thickened torus $T^2 \times I$ (Fig~\ref{criteria}), which can then lead to a link $L'_{\rm c}$ in $\mathbb{R}^3$. 
The same process applies to the grid $G_{\rm o}^n$ which includes defect cells, generating a link $L_{\rm o}$ in $T^2 \times I$ (Fig~\ref{criteria}(a)), and the corresponding link $L'_{\rm o}$ in $\mathbb{R}^3$. The knitting Criterion I is defined as follows:

\smallbreak
\textit{A textile pattern generated by the unit cell $A$ is defined as knittable according to Criterion I if $L'_{\rm o}$ is a trivial link in $\mathbb{R}^3$ while $L'_{\rm c}$ is not.}
\smallbreak

Knitting Criterion II --- 
As in the case of Criterion I, we consider the torus diagram resulting from the gluing of the opposite sides of the grids $G_{\rm c}$ and $G_{\rm o}^n$. 
In particular, the gluing of the left-right sides of the grids forms a meridian of the torus.
We now twist the torus along this meridian, clockwise or counterclockwise, such that each row of the grid is glued to its adjacent upper row on one side of the meridian, and to its adjacent bottom row on the other side of the meridian, as illustrated in Fig~\ref{criteria}(b). 
This shifts the recombination along the columns of the grids by one row so that the cell $(i,l-1)$ is connected to the cell $(i-1,0)$. 
The resulting objects are the knots $K_{\rm c}$ and $K_{\rm o}$ in the thickened torus, resulting from the grids $G_{\rm c}$ and $G_{\rm o}^n$, respectively, after twisting the torus diagrams.
These knots are made of a single curve following a helicoid trajectory winding around the torus.
Then, by mapping $K_{\rm c}$ and $K_{\rm o}$ in the three-space as done in Criterion I, we obtain the knots $K'_{\rm c}$ and $K'_{\rm o}$ in $\mathbb{R}^3$.
The Knitting Criterion II is defined as follows :

\smallbreak
\textit{A textile pattern generated by the unit cell $A$ is defined as knittable according to Criterion II if $K'_{\rm o}$ is a trivial knot in $\mathbb{R}^3$ while $K'_{\rm c}$ is not.}
\smallbreak

For a given textile pattern, these criteria provide a topology-based definition of knittability. They also provide a classification method, whether a knittable textile satisfies one or two of the criteria. 
If a pattern satisfies neither criterion, it is not knittable, within the constraints we imposed on the pattern (Fig.~\ref{venn}). 
If a textile satisfies the Knitting Criterion I, we include it in the \textit{knits} category, but if it satisfies only the 
Knitting Criterion II, we classify it in the `\textit{crochet} category. 
Note that the topological modeling of crochet based on the twist operation of the torus, as presented in this work, is an original construction. 
In the next section, we will discuss the implementation of these criteria and the rationale behind the classification.

\begin{figure*}
    \centering
    \includegraphics[width=17cm, bb = 0 0 167 165]{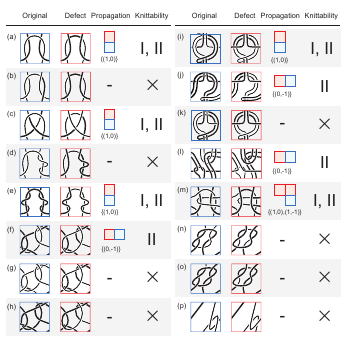}
    \caption{Application of the knittability criteria to a range of textile patterns. For each example (a-p), the unit cell, the corresponding defect, the propagation set, and the criteria satisfaction are featured. As they satisfy at least one criterion, the patterns (a, c, e, f, i, j, l, m) are considered knittable.
    }
    \label{examples}
\end{figure*}

\section{Criteria implementation}

We have reduced the determination of knittability to estimating whether knots and links are trivial or not. 
In knot theory, the classification and the unknotting problems are long-standing goals, which motivated the developments of computational tools~\cite{haken1961theorie,hass1999computational,khovanov2000categorification,bar2005khovanov,kronheimer2011khovanov}.
However, due to the high computational cost of the algorithm (typically growing exponentially with the number of crossings), it is challenging to find a systematic solution to this problem. 
A simple heuristic method is therefore a reasonable and practical alternative. 
Here, the concept of defect propagation introduced earlier (Fig.~\ref{copy}) comes into play. 

For a given textile diagram, we aim at characterizing whether $K'_{\rm o}$ and $L'_{\rm o}$ are trivial.
Recall from the previous section that these knots and links are intuitively obtained by folding on the torus the grid $G_{\rm o}$ into which a row of unit cells $A$ has been replaced by its defect equivalent $\tilde A$ that disentangle the yarns along an effective working edge. 
We established that these defects can propagate in various directions (Fig.~\ref{copy}), thus, the initial row of defects may invade the whole torus depending on its propagation set. 
For $L_{\rm o}$ and the Knittability Criterion I, the row of defect loops back on itself, so it invades the whole torus if propagation has only components along a column, perpendicular to the yarn direction as in Fig.~\ref{copy}(a). 
In term of propagation sets, the first coordinate of each position must be non-zero, with all the same sign, such as $\{(1, -1),\,(1, 0),\,(1, 1)\}$ or $\{(-1, -1),\,(-1, 0)\}$.
A propagation along the yarn direction (along a row in the grid, as for Fig.~\ref{copy}(b)), keeps the defect on the same row. 
For $K_{\rm o}$ and the Knittability Criterion II, the rows of defects loop back on the row below.
In that case, defect propagation along the yarn direction also invades the whole torus by spiraling along the shifted rows. 
This time, the propagation sets featuring zero or different components in the first coordinate of each position also invade the whole torus.

Depending on the propagation set of the defects, we can establish that $K'_{\rm o}$ or $L'_{\rm o}$ is equivalent to a link made from a grid containing only defects $\tilde A$. 
Yet, defects disentangle the yarn, so any link made exclusively from $\tilde A$ must be trivial.
The propagation sets provide a way to show the existence of a sequence of Reidemeister moves that drives the number of crossings in the diagram of $K'_{\rm o}$ or $L'_{\rm o}$ to $0$. 
Evidence that $K'_{\rm o}$ or $L'_{\rm o}$ is trivial is given by the propagation set of the defects. 
The problem of knittability is then shifted from characterizing a complex link to the heuristic determination of propagation sets.
Now, we have to provide evidence that the knots and links without defects, $K'_{\rm c}$ and $L'_{\rm c}$, are in contrast non-trivial. 
If $K'_{\rm c}$ and $L'_{\rm c}$ were trivial, it means that the yarns would not entangle upon constructing the fabric, hence at the unit cell level. 
The unit cells of such non-entangling textiles would then be equivalent to their defect analogues. 
If a unit cell $A$ is equivalent to its defect $\tilde A$ by Reidemeister moves within the unit cell, then the textile is not knittable as it would form a fully disentangled structure.

To summarize, the implementation of the knittability criteria is as follows, for a given textile diagram made of unit cell $A$:
\begin{itemize}
    \item Determine the defect $\tilde A$ following the process detailed in Sec.~\ref{defect_propagation}.
    If $A$ and $\tilde A$ are topologically equivalent, the pattern is unknittable; if not, we assess the propagation sets of the defect.
    \item If defects propagate along a column (perpendicular to the yarn direction), then both $K'_{\rm o}$ and $L'_{\rm o}$ are equivalent to trivial links and both criteria are satisfied. If the defects propagate along a row (the direction of the yarn), only the Knittability Criterion II is satisfied. In both cases, the pattern is deemed knittable. 
    \item If defects do not propagate, none of the Knittability Criteria are satisfied, the pattern is deemed unknittable under our definition. 
\end{itemize}

We implemented this approach over a wide variety of patterns (Fig.~\ref{examples}) and successfully distinguished known knittable patterns from unknittable ones. 
For knittable patterns, satisfaction of one or two of the knittability criteria enables a distinction between two subcategories. 
Patterns whose defect propagation set has only vertical components satisfy both knittability criteria, and we classify them as \textit{knits}. 
Those whose propagation set has a horizontal component satisfy only the Knittability Criterion II and are coined as \textit{crochet}. 
The set of \textit{knits} comprises all the patterns classically associated with weft knitting, while the \textit{crochet} categories encompass patterns used in crochet and warp knitting. 
The fabrication processes associated with these two subcategories differ greatly, so beyond knittability, details on the fabrication techniques are also stored in the topological properties of the patterns.
For instance, the patterns of Fig.~\ref{examples}(a) and (f) represent, respectively, a standard knitted and crocheted unit cell. 
The knit unit (a) propagates vertically following a propagation set $\{(1,0)\}$, hence an entire line of defects in $G_{\rm c}$ and $G_{\rm o}$ is necessary to propagate the defects everywhere on the torii. 
This defect line reflects the presence of the free working edge during the fabrication of knits. 
In contrast, the crocheted unit in Fig.~\ref{examples}(f) propagates laterally according to the set $\{(-1,0)\}$, then remarkably, introducing only a single defect in $G_{\rm o}$ is sufficient to untangle completely $K'_{\rm o}$. 
The structural fragility to a single defect reflects that crocheting progresses loop by loop, with the previous one topologically secured, reducing the working edge to a single loop.

This method appears efficient in characterizing knittability if we identify, by a succession of Reidemeister moves, the propagation sets of a defect. 
However, finding a propagation set is not a trivial task, and we cannot prove its absence.
Therefore, unknittability cannot be proven with the method described. 
We now aim to confirm our results, obtained from propagation, with knot theory standard tools. 
Since our knittability criteria are evaluated on the triviality of $K'_{\rm o}$ and $L'_{\rm o}$, an ideal tool would detect triviality unambiguously. 
In knot theory, a topological invariant is any quantity assigned to a link that remains unchanged under equivalence. 
In particular, all trivial links with the same number of components necessarily share identical values of any given invariant.
However, the converse fails in general: no known invariant is both necessary and sufficient for triviality, so that some non-trivial links may nonetheless share the same invariant values as trivial ones. 
The only widely studied exception is Khovanov homology, a tool that assigns algebraic structures to a knot in such a way that only the trivial knot produces the simplest possible result~\cite{kronheimer2011khovanov}. 
While Khovanov homology can detect the unknot, its high computational cost for moderately complex knots and its inability to detect trivial links motivate the search for a more direct, systematic approach. 
We therefore adopt a brute-force strategy of computing simpler invariants for each pattern’s associated links $K'_{\rm o}$ and $L'_{\rm o}$.

\begin{figure}
    \centering
    \includegraphics[width=8.6cm, bb = 0 0 156 188]{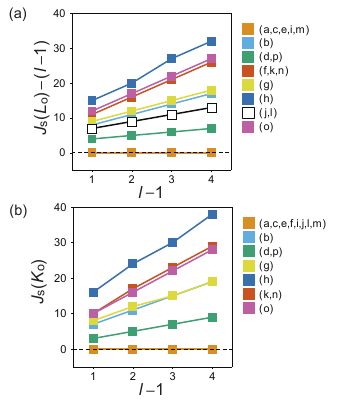}
    \caption{(a) The span of the Jones polynomial $J_{\rm s}$ of the knot $K_{\rm o}$ is plotted against the number of passes $l$ for all the patterns shown in Fig.~\ref{examples}. Patterns with $J_{\rm s}(K_{\rm o})=0$ satisfy criterion I. (b) For the same patterns, the quantity $J_{\rm s}-(l-1)$ for the links $L_{\rm o}$ is plotted against $l$. Patterns satisfying criteria 2 must yield $J_{\rm s}(L_{\rm o})-(l-1)=0$. These computations provide a quantitative validation of the knittability criteria estimated via the propagation set.
    }
    \label{plot}
\end{figure}

Rather than relying on completeness, one can exploit the ability of an invariant to measure complexity. 
We use the span of the Jones polynomial, denoted $J_{\rm s}$, which is defined as the difference between the highest and lowest degrees in the Jones polynomial $V_L(t)$~\cite{jones1997polynomial}. 
The Jones polynomial is widely used in knot theory due to its computability and its ability to distinguish between different knot and link types.
Its span $J_{\rm s}$ provides a single scalar summarizing entanglement complexity, and takes especially simple values on trivial links: for the unknot, $J_{\rm s}=0$, and for a trivial link of $l$ components, $J_{\rm s}=l-1$. 
Moreover, computing $J_{\rm s}$ remains feasible for unit cells with a moderate number of crossings. We therefore calculate the span of $V_{K'_{\rm o}}(t)$ and $V_{L'_{\rm o}}(t)$ using the knotfolio library~\cite{knotfolio}, applied to textile diagrams with two columns ($m=2$) and a varying number of rows $l$ (Fig.~\ref{plot}). 
Our results mirror those from the propagation method: knittable patterns under the Knittability Criterion I yield $J_{\rm s}(K'_{\rm o})=0$, and those under the Knittability Criterion II yield $J_{\rm s}(L'_{\rm o})=l-1$ for all $l$. 
In contrast, unknittable patterns exhibit higher and increasingly diverse spans as $l$ grows, reflecting the greater topological complexity of their associated links. 
Combined with the propagation approach, this span-based method thus offers a practical conjecture for the knittability or unknittability of textile patterns whose unit cells have manageable crossing numbers.

Akin to reducing a many-body to a one-body problem, we leveraged periodicity and local propagation sets to reduce the complexity of knots and links. 
It results in a straightforward implementation to determine if a textile pattern is knittable. 
We validated the approach by directly confronting with the estimation of knot invariants. 
In the next section, we discuss further physical analogies and potential applications of the relation between topology, knittability, and defect propagation.

\section{Discussion}

\subsection{Amimono as marginally constrained textiles}

The criteria we implemented to define amimono fabrics highlight a remarkable property: knits and crochets are marginally constrained textiles. 
Knitted fabrics are stable structures, but the introduction of some defects completely destroys the structure. One can draw an analogy with other marginally constrained systems whose stability is assessed by constraint counting. 
Counting the number of degrees of freedom and constraints has been instrumental in identifying floppy modes of metamaterials, such as frames~\cite{paulose2015topological} or origami~\cite{chen2016topological}, but also to locate rigidity transitions in many-body systems ~\cite{ohern2003jamming,yan2019multicellular}. 
With $f$ the number of degrees of freedom of a system, $c$ its number of constraints, we define their difference by $\nu=f-c$. Over-constrained ``rigid'' systems have $\nu<0$, while under-constrained ``floppy'' ones have $\nu>0$. 
The isostatic point at $\nu=0$ marks the rigidity transition.
By analogy, we define for a patch of textile $\nu=-\tilde N/ N$, the ratio between the minimal number of defects to introduce to disentangle the patch, $\tilde N$, and the number of unit cells, $N=l\times m$. As the number of cells increases, the number of crossings also increases. 

In this case, the increase in the number of crossings is not due to Reidemeister moves but by increasing $l$ and $m$ only. 
Thus, since the number of unknotting operations is fixed in the unit cell, increasing $l$ and $m$ necessarily increase the number of crossings to be switched, and inversely, adding defects decreases the number of crossings. 
The absolute value of the ratio $|\nu|$ is thus akin to the average of the unknotting number per unit cell.
For unknittable textiles, as defects do not propagate, the number of defects necessary equals the size of the textile $\tilde N = N$. Unknittable textiles have then $\nu=-1$, which would classify them as over-constrained systems. 
In contrast, defects in amimono propagate and invade lines or columns. 
Therefore, the number of defects necessary to introduce is systematically smaller than the size of the textile patch. 
Even if the value of $\tilde N$ depends on the method of folding the patch on itself, the asymptotic value of $\nu$ will systematically converge towards the isostatic limit $\nu=0$. 
Amimono are then characterized by being marginally constrained textiles in the asymptotic limit of large fabric size.

The marginal stability of amimono has direct practical implications. 
As a method of identification, it suffices to remove or cut a yarn in a textile and observe the propagation of the damage to assess its method of fabrication. 
If the damage easily propagates, the textile has probably been manufactured using local manipulations of the yarn, as in amimono. 
When such a destructive method is not adapted, as for textiles excavated from archaeological sites~\cite{good2001archaeological,obata2025nets} or synthesized at the molecular scale~\cite{forgan2011chemical,august2020self}, we can extract the topological structure and virtually place defects to infer their propagation. 

\subsection{Defect approach to functional textiles}

\begin{figure}
    \centering
    \includegraphics[width=8.6cm, bb = 0 0 173 204]{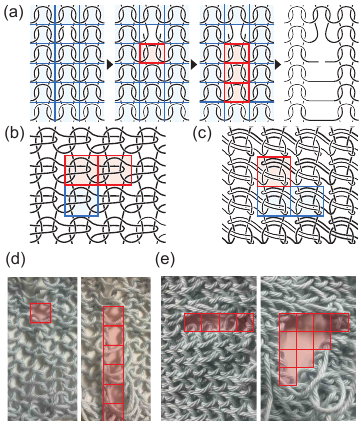}
    \caption{
    (a) Defect propagation from a cut yarn, leading to the laddering effect of knitted fabrics. 
    (b) Robust knit, designed such that two defects on the upper row are needed to propagate one defect on the row below. 
    (c) Fragile knit, one defect propagates into two defects on the row below. 
    (d) Laddering of the widely used jersey knit pattern. 
    (e) Realization of the robust knit, where four defects were initially placed on one row. When propagating to a subsequent row, the number of defects is reduced by one, limiting the topological fracture to 4 rows and 10 defects in total.
    }
    \label{ladder}
\end{figure}

Another direct implication is the rationalization of a very well-known and troublesome feature of knits, laddering (Fig.~\ref{ladder}(a)). 
When a loop is missed during manufacturing or if the yarn is damaged and locally cut, a ``ladder'' rapidly forms in a knitted textile under tension. 
Laddering is a direct visualization of topological defect propagation in amimono, with a breakdown of the structure over a long range, even if caused by a local perturbation. 
A remarkable property of laddering is that the nature of the propagated damage is topological and not material. 
Apart from the initial cut, the thread remains intact. 
As loops get freed by the propagation of the topological defect, the tension in the fabric straightens the freed loops, making the ladder apparent without any further damage and even minimal deformation to the threads. 
The direction of propagation is also completely determined by topology, and not by the direction of the yarn. 
Laddering can then be classified as a topological fracture, in contrast with classic fracture induced by irreversible breaking of materials. 
Since the propagation of the topological fracture is driven by local and reversible deformation of the yarn, they are completely reversible, except for the initial damage. 
All knitters have likely experienced a slipping loop that propagates by laddering, but we also know that it does not mean the knit is ruined: by carefully looping back the ladder, we impose a sequence of Reidemeister moves that heal the topological fracture. 

The criteria established to determine whether a textile can be knitted or not, provide a topological basis to design functional textiles. 
In particular, one can use the analysis of defect propagation to bring specific properties and functions to fabrics. 
As an example, we propose a geometry that mitigates the problem of laddering. 
The two patterns in Fig.~\ref{ladder}(b,c) are both knittable textiles with radically different resistance to laddering embedded into their structure. 
The textile in Fig.~\ref{ladder}(b) restrains the propagation of laddering via its defect propagation sets. The propagation, expressed as $\mathcal{S}=\{(0,1),(1,1)\}$, is such that from $n$ adjoint defects along a yarn, the yarn below will have $n-1$ defects. 
This structure forces the number of defects to shrink as they propagate, limiting the laddering phenomena that can propagate all across the fabric for standard jersey kitting, for instance (Fig.~\ref{ladder}(a)).
In contrast, the textile in Fig.~\ref{ladder}(c) is designed to be extremely sensitive to defects. 
This time, the propagation set given by $\mathcal{S}=\{\{(-1,1)\},\{(0,1)\}\}$ features two directions of propagation. 
The structure of the resulting fabric is then entirely destroyed with a single defect.

The possible operations in amimono closely resemble the dynamics of kinetically constrained models (KCM)~\cite{garrahan2011kinetically}. As discussed in Sec.~\ref{defect_propagation}, whether a given cell in an amimono structure can be transformed between the states $A$ and $\tilde A$ depends on the configuration of neighboring cells. 
This constraint is analogous to a defining feature of KCM, where the possibility of spin flipping at a site is determined by the local spin configuration. 
This analogy provides a direct correspondence between the propagation sets characterizing defect propagation in amimono and the dynamical constraints in KCM. 
For instance, the propagation set $\{(1,0)\}$ studied in this work corresponds directly to the East model\cite{jackle1991hierarchically,sollich1999glassy}, a well-known example within the KCM class. 
This analogy suggests that tools developed for KCM may be applicable to amimono, particularly in analyzing the asymmetry between knitting and unknitting operations, as well as the stability of randomly generated knit structures.

\section{Conclusion}

Textiles come in many topologies, reflecting the various crafting techniques developed across human history. 
In this work, we established a framework based on knot theory to create a direct link between the topology of a textile and its fabrication method. 
Knits and crocheted works are obtained by manipulating a yarn with fixed ends, a feature that remains encoded in the topology of the fabric and that we leverage in a method to assess knittability. 
We start by considering doubly periodic textile patterns and rule out obvious non-knittable patterns. 
The resulting textile patterns can be described as a matrix of repeated unit cells, with yarns running in the direction of the rows. 
We then introduce the concept of topological defects in textile diagrams by defining the least untangled alter ego of a unit cell. 
Depending on the topology of the textile, the topological defects may propagate to neighboring cells by transforming them into defects without cutting the yarns.

Subsequently, knittability criteria are established by considering a swatch of unit cells, introducing defects, folding the swatch into a torus, and finally assessing the triviality of the resulting knot or link. 
The crafting constraints of amimonos, with only local manipulation of loops, make their structure sensitive to topological defects, which, in a sense, locally invert the crafting process. 
Hence, if defects can propagate, they will cover the entire torus and disentangle the yarns. 
A textile is then knittable if its defect-infused diagram folded onto a torus results in a trivial knot or link. 
Furthermore, we introduce two criteria to distinguish whether defects propagate along columns or rows by adding a slight twist to the torus.
The two criteria enable refined classification between \textit{knit} and \textit{crochet} textiles, which share the local manipulation feature but differ in the loop entanglement method. 

This work opens both fundamental questions and practical applications. 
The assessment of defect propagation is currently performed heuristically, but using geometric or combinatorial methods to automatically establish the propagation patterns would unlock the possibility of dealing with arbitrarily complex textiles with large unit cells. 
Besides, the pattern of defects infused into the swatch is chosen such that standard amimono diagrams are deemed knittable. 
Yet, an interesting direction lies in determining the minimal set of defects to verify a criterion, allowing finer classification and extraction of deep topological features of textiles. 
We also demonstrated that the breakdown of a textile is tunable by designing the defect propagation pattern, paving the road for failure-aware functional fabrics. 
Ultimately, the framework we built reveals surprising analogies between textiles and other physical systems, including marginally constrained structures and kinetically constrained models. 
Exploring these relations further may help us understand how the topology influences the physics and mechanics of entangled materials with, beyond textiles, applications in soft robotics, polymer chemistry, and biological macromolecules.

\begin{figure*}[]
    \centering
    \includegraphics[width=14cm, bb = 0 0 251 154]{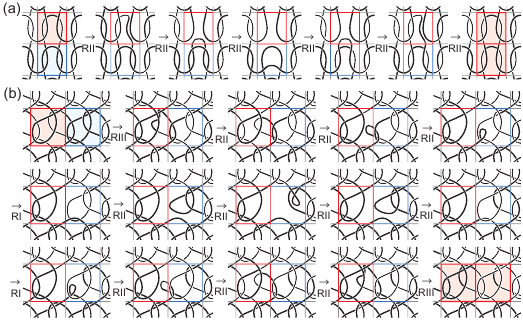}
    \caption{(a,b) The sequences of Reidemeister moves to propagate the defects in examples in Fig.~\ref{copy} (b) and (c), respectively.
    }
    \label{Annexe_reidemeister}
\end{figure*}

\begin{acknowledgments}
This work benefited from fruitful discussions with, but not limited to, K. Masuda, T. Hatano, R. Suzuki, Y. Takaha, H. Katsuragi, T. Yamaguchi and Y. Uesugi. 
SD acknowledges the financial support provided by JSPS KAKENHI Grant-in-Aid for JSPS Fellows 23KJ0753. 
SM acknowledges the support by RIKEN iTHEMS and by the JSPS KAKENHI Grant-in-Aid for Early-Career Scientists 25K17246. SP acknowledges the financial support provided by JSPS KAKENHI Grant-in-Aid for Early-Career Scientists 25K17363. 
\end{acknowledgments}

\appendix*

\section{Reidemester moves sequences for defect propagation}
For the pattern examples shown in Fig.~\ref{copy}(b-c), we demonstrate the sequences of Reidemester moves that constitute their propagation pattern (Fig.~\ref{Annexe_reidemeister}).

\newpage

\bibliography{knit}

\end{document}